\documentclass[journal=cmatex,manuscript=article]{achemso}

\usepackage[version=3]{mhchem} 
\usepackage{amsthm}
\usepackage{amssymb}
\usepackage{amsfonts}
\usepackage[dvipsnames]{xcolor}

\author{F\'elix~Thouin}
\affiliation[GT-phys]{School of Physics, Georgia Institute of Technology, 837 State Street NW, Atlanta, Georgia 30332, United~States}
\altaffiliation{Contributed equally to this work}

\author{Ajay~Ram~Srimath~Kandada}
\affiliation[IIT]{Center for Nano Science and Technology@PoliMi, Istituto Italiano di Tecnologia, via Giovanni Pascoli 70/3, 20133 Milano, Italy}
\altaffiliation{Contributed equally to this work}

\author{David~A.~Valverde-Ch\'avez}
\affiliation[GT-chem]{School of Chemistry and Biochemistry, Georgia Institute of Technology, 901 Atlantic Drive NW, Atlanta, Georgia 30332, United~States}

\author{Daniele~Cortecchia}
\affiliation[IIT]{Center for Nano Science and Technology@PoliMi, Istituto Italiano di Tecnologia, via Giovanni Pascoli 70/3, 20133 Milano, Italy}

\author{Ilaria~Bargigia}
\affiliation[GT-chem]{School of Chemistry and Biochemistry, Georgia Institute of Technology, 901 Atlantic Drive NW, Atlanta, Georgia 30332, United~States}

\author{Annamaria~Petrozza}
\affiliation[IIT]{Center for Nano Science and Technology@PoliMi, Istituto Italiano di Tecnologia, via Giovanni Pascoli 70/3, 20133 Milano, Italy}

\author{Xunmo~Yang}
\affiliation[Google]{Google, 1600 Amphitheatre Parkway, Mountain View, California 94043, United~States}

\author{Eric~R.~Bittner}
\affiliation[UH]{Department of Chemistry, University of Houston, Houston, Texas 77204, United~States}
\alsoaffiliation[Durham]{Department of Physics, Durham University, South Road, Durham, DH1~3LE, United~Kingdom}
\email{ebittner@central.uh.edu}

\author{Carlos~Silva}
\affiliation[GT-chem]{School of Chemistry and Biochemistry, Georgia Institute of Technology, 901 Atlantic Drive NW, Atlanta, Georgia 30332, United~States}
\alsoaffiliation[GT-phys]{School of Physics, Georgia Institute of Technology, 837 State Street NW, Atlanta, Georgia 30332, United~States}
\email{carlos.silva@gatech.edu}

\title[Exciton polaron dynamics]
  {Electron-phonon couplings inherent in polarons drive exciton dynamics in two-dimensional metal-halide perovskites\footnote{This article is submitted to the Jean-Luc Br\'edas Festschrift}}

\begin{document}







\newpage
\begin{abstract}
  We report on the exciton formation and relaxation dynamics following photocarrier injection in a single-layer two-dimensional lead-iodide perovskite. We probe the time evolution of four distinct exciton resonances by means of time-resolved photoluminescence and transient absorption spectroscopies, and find that at 5\,K a subset of excitons form on a $\lesssim$ 1-ps timescale, and that these relax subsequently to lower-energy excitons on $\sim$ 5--10\,ps with a marked temperature dependence over $<$ 100\,K. We implement a mode projection analysis that determines the relative contribution of all observed phonons with frequency $\leq$50\,cm$^{-1}$ to inter-exciton nonadiabatic coupling, which in turn determines the rate of exciton relaxation. This analysis ranks the relative contribution of the phonons that participate in polaronic lattice distortions to the exciton inter-conversion dynamics and thus establishes their role in the nonadiabatic mixing of exciton states, and this in the exciton relaxation rate. 
\end{abstract}

\section{Introduction}
Hybrid organic-inorganic metal-halide two-dimensional perovskites (2D-HOIPs) have attracted considerable interest over the past two decades due to their promising optoelectronic characteristics~\cite{grancini2018dimensional, chen20182d} and due to their structural and dynamic complexity as strongly quantum-confined materials~\cite{smith2019tuning, quintero2018compositional}. These are direct-bandgap semiconductors that are composed of two-dimensional (2D) planes of ionic metal-halide lattice layer separated by long organic cations. Thus, carriers are confined to the few-atom-thick inorganic lattice planes. Due to strong quantum~\cite{blancon_scaling_2018} and dielectric~\cite{Hong1992} confinement effects, tightly bound excitons and biexcitons~\cite{Ishihara1992,thouin_stable_2018,neutzner_exciton-polaron_2018, even2014understanding} are primary photoexcitations. The ionic character and the softness of the lattice lead to strong electron-phonon interactions~\cite{zhang2018rapid} while the localized uncorrelated motion of the organic cations leads to strong dynamic disorder~\cite{thouin_stable_2018}. Unlike other 2D semiconductors such as III-V quantum wells and transition-metal dichalcogenide monolayers, very clear spectral structure is present in the exciton absorption in this class of materials, which reflects the presence of at least four distinct 
resonances that are evenly separated in energy~\cite{kataoka_magneto-optical_1993}, which are spectrally correlated as revealed by coherences via a common ground state and via biexcitons in coherent two-dimensional spectra~\cite{thouin_stable_2018,neutzner_exciton-polaron_2018}, establishing that the spectral structure is intrinsic. 
We have recently demonstrated by analysis of resonant impulsive stimulated Raman coherences that these multiple excitons and photo-carriers are dressed by lattice phonons via polaronic interactions~\cite{thouin_phonon_2019}, which indicates that the multiple resonances represent distinct excitons as opposed to a vibronic progression of a single exciton. We also observed that the lattice configurations of the excited states (free carriers and multiple excitons) are displaced along specific and distinct lattice coordinates, predominantly involving motion within the two-dimensional lead iodide layer. This suggests a complex scenario for carrier relaxation and exciton formation where excitation dynamics may be driven by specific lattice phonons that also participate in the lattice polaronic reconfiguration. Here, we address the role of the phonon modes that dress the exciton-polarons in the exciton relaxation dynamics and distinguish between active \emph{driving} modes from passive \emph{spectator} modes.

We employ temperature-dependent transient absorption and time-resolved photoluminescence spectroscopies on (PEA)$_2$PbI$_4$ (PEA = phenylethylammonium), a prototypical 2D-HOIP. We use a mode projection scheme~\cite{pereverzev2009energy,yang2014intramolecular,yang2015computing,Yang:2017aa}, in which experimentally measured phonon frequencies and Huang-Rhys parameters~\cite{thouin_phonon_2019} are used to construct the smallest possible subspace of modes that optimizes the electron-phonon coupling that drives exciton dynamics. We find that a subset of three modes is sufficient to reproduce the experimentally measured electron-phonon coupling parameters, and we link these to the formation dynamics of the lowest-energy exciton on picosecond timescales. We, therefore, conclude that these specific modes drive the exciton formation dynamics, with the rest of the polaronic dressing modes predominantly playing a spectator role. With this analysis, we highlight the correlation between the phonon dynamics associated with polarons and the photoexcitation relaxation dynamics following hot carrier injection, which is of fundamental importance in light-emitting diodes and related optoelectronic technologies.

\section{Results and analysis}

\subsection{Linear spectroscopy}

\begin{figure}[tbh]
\centering
\includegraphics[width=0.5\columnwidth]{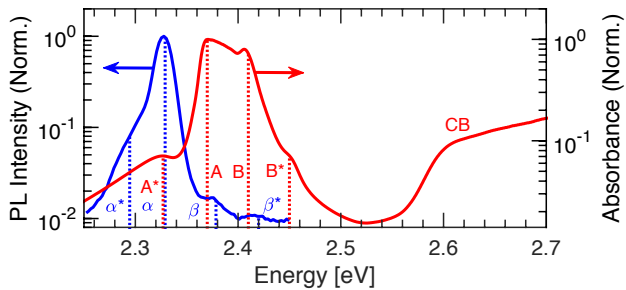}  
\caption{Normalized absorption (red) and time-integrated photoluminescence (PL) intensity (blue) of a polycrystalline film of single-layer (PEA)$_2$PbI$_4$ at a temperature of 5\,K. Features in the PL spectrum are labelled by blue Greek characters while those in the absorption spectrum are labelled by red Latin characters. Their spectral positions are denoted by dotted lines of the appropriate color. We note that the PL spectrum is not corrected for the experimental spectral response.}
\label{fig:abs_and_pl}
\end{figure}

The absorption and time-integrated photoluminescence (PL) spectra of a polycrystalline film of single-layer (PEA)$_2$PbI$_4$ at a temperature of 5\,K are presented in Fig.~\ref{fig:abs_and_pl}. We have reported the complete structural and optical characterization of this material elsewhere~\cite{thouin_stable_2018,neutzner_exciton-polaron_2018}. Four distinct transitions (labelled A*, A, B and B*) are observed in the absorption spectrum, with the most intense resonance (A) peaked at 2.36\,eV, which is 253\,meV below a sharp continuum band edge (labelled CB)~\cite{neutzner_exciton-polaron_2018}. Four features (labelled $\alpha$*, $\alpha$, $\beta$ and $\beta$*) are also observed in the PL spectrum, and are red-shifted by approximately 35\,meV from their associated peaks in the absorption spectrum. The relative intensity of these peaks differs from those observed in the absorption spectrum, indicating inter-species relaxation before the population of the final emissive states. To probe these exciton dynamics, we performed transient absorption (TA) and time-resolved photoluminescence (TRPL) spectroscopies.

\subsection{Exciton formation and decay dynamics}

\begin{figure}[tbh]
\centering
\includegraphics[width=0.7\columnwidth]{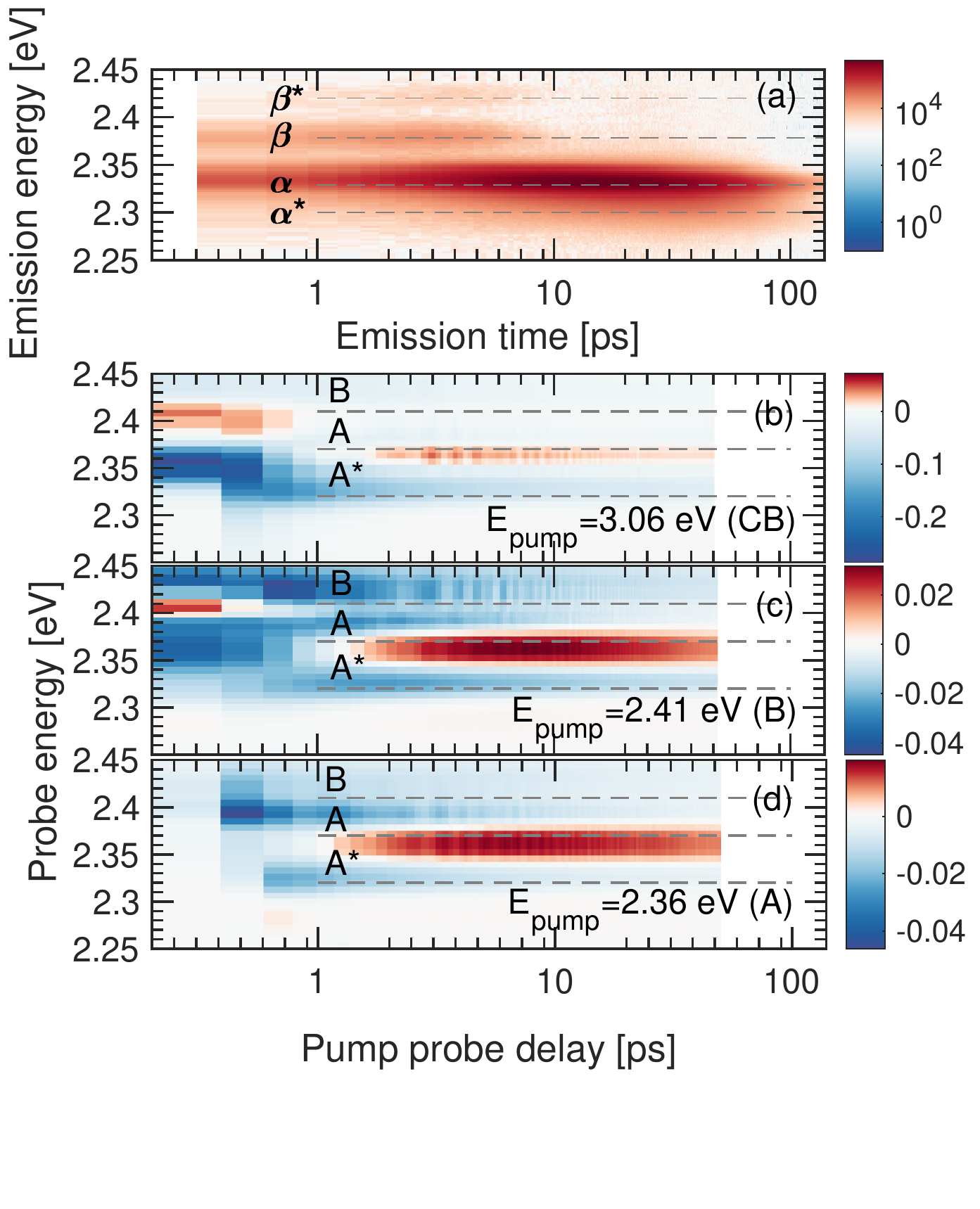}
\caption{(a) Time-resolved photoluminescence intensity of the sample at 5\,K under low excitation fluence. The features identified in Fig. \ref{fig:abs_and_pl} are indicated by dotted lines and labelled accordingly. Time-resolved differential transmission ($\Delta T/T$) spectra measured at 5\,K at a constant pump photon flux of ($4.32\times 10^{11}$ photons/cm$^2$) when pumping (a) hot carriers into the CB, (c) exciton B,  or (d) exciton A.}
\label{fig:pl_and_tals}
\end{figure}

We first focus on dynamics observed in TRPL at a temperature of 5\,K upon injection of hot carriers by tuning the pump energy to 3.06\,eV, displayed in Fig.~\ref{fig:pl_and_tals}(a). The four transitions identified in the time-integrated PL spectrum are also observed, but with contrasting temporal evolution. Within the picosecond time resolution of our instrument, we observe prompt formation of features $\beta$ and $\beta$*, and they decay on a longer picosecond timescale, concomitant with a rise of feature $\alpha$, followed by feature $\alpha$* (refer to Figure S1 in the Supplemental Information for the dynamics). We note that this TRPL behavior is consistent with that reported previously~\cite{straus_direct_2016}.

We observe related dynamics in corresponding TA measurements upon pumping at 3.06\,eV and probing the exciton bleach spectral region. The differential transmission ($\Delta T/T$) spectrum under these conditions is shown in Fig.~\ref{fig:pl_and_tals}(b). In general, the transient bleach lineshape in HOIPs is dominated by carrier thermalization~\cite{wu2015trap, giovanni2018coherent}, exciton-screening mechanisms~\cite{schmitt1985theory}, and photoinduced changes of the permittivity function~\cite{price2015hot}, so we refrain from mapping the bleach lineshape onto the linear spectrum. We note, however, that a positive feature with instrument-limited rise time at a probe photon energy of 2.45\,eV decays on sub-picosecond timescales, concomitant with the evolution of an initially negative feature at a probe energy of 2.36\,eV into a longer-lived positive component. We also highlight a negative differential transmission feature at a probe energy of 2.32\,eV, which rises within hundreds of femtoseconds and monotonously decays over tens of picoseconds. We note the similarity in the timescale of the transient behavior of the feature at 2.36\,eV in the TA spectrum with the rise time of the $\alpha$ peak in Fig~\ref{fig:pl_and_tals}(a). We also associate the decay of the 2.32-eV negative signal in Fig.~\ref{fig:pl_and_tals}(b) with that of the $\beta$ peak in Fig.~\ref{fig:pl_and_tals}(a) (see also Fig. S3 of the Supplemental Information). 

We examine further the TA spectral features by resonantly exciting the two main excitonic transitions, namely pumping exciton B or A. These TA spectra are presented in Fig.~\ref{fig:pl_and_tals}(c) and (d), respectively. Upon pumping exciton B, we observe similar temporal evolution at 2.36\,eV, namely an initially negative signal that evolves into a positive signal with a time constant of several picoseconds. On the other hand, when we pump exciton A resonantly, we only observe the corresponding rise of a positive signal at that probe energy. This supports the conjecture that the dynamics at this probe energy represent the B $\to$ A exciton conversion, and can be associated with the dynamics of the $\beta$ peak in TRPL. We also note that the relaxation dynamics probing at 2.32\,eV are weakly dependent on whether we excite into the conduction band or either exciton. 

\begin{figure}[tbh]
\centering
\includegraphics[width=0.5\columnwidth]{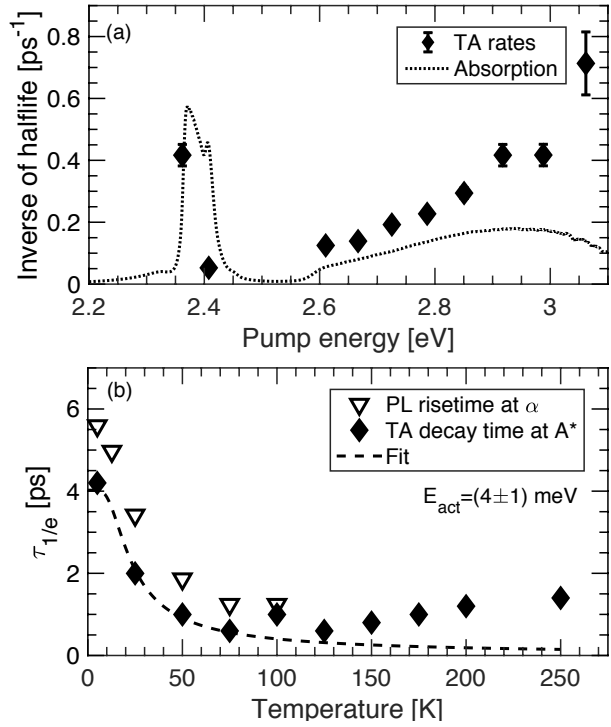}
\caption{(a) Inverted half-life of the 2.32-eV TA trace (filled diamonds) compared with the absorption spectrum (dotted line). The uncertainties represent the time resolution of our measurements (200\,fs). (b) Dependence of the lifetime of the 2.32-eV TA feature on sample temperature at a pump energy of 3.06\,eV. Plotted alongside is the rise time of the $\alpha$ TRPL feature when pumped at the same energy. The dotted line represent the best fit of the rate model described in the text.}
\label{fig:rates}
\end{figure}

The picture that emerges from Fig.~\ref{fig:pl_and_tals} is that upon injection of hot photocarriers at a temperature of 5\,K, these recombine to populate excitons B and B* within $\sim 1$\,ps. These excitons feed excitons A and A* over several picoseconds, and the resulting population decays on tens of picoseconds. We next consider the decay dynamics of the transient feature at 2.32\,eV given its omnipresence at all three pump energies. Given non-exponential decay behavior, we plot the inverse half-life as a function of pump energy. The decay-rate dependence on pump energy at 5\,K is presented in Fig.~\ref{fig:rates}(a), along with the low-temperature absorption spectrum for comparison. The decay rates increase monotonously with increasing excess energy above the continuum band edge. This pump-energy dependence is consistent with carrier cooling dynamics driven by phonon scattering. We highlight that even at this low temperature, the relaxation time ranges from $\sim 1.5$\,ps at the highest probed energy to $\sim 20$\,ps near the bandedge. Such rapid relaxation dynamics are unsurprising in a semiconductor in which electron-phonon coupling is strong. Strikingly, when pumping resonantly exciton B, the decay rate is about an order of magnitude lower than that upon resonant pumping of exciton A. This marked pump excitation spectrum reveals that the exciton decay rate depends strongly on the nature of the vertical excitation: pumping into the conduction band reflects a pump energy dependence that is consistent with hot carrier relaxation~\cite{giovanni_tunable_2016, hunsche1993spectral}, while excitons that are pumped resonantly display clear spectral structure on the decay rate. Given the resonant exciton-phonon coupling excitation spectrum reported in Ref.~\citenum{thouin_phonon_2019}, this implies that excitons are locked depending on the polaronic nature of the vertical excitation.

The temperature dependence of the $1/e$ decay time of the same 2.32-eV TA feature upon 3.06-eV excitation is plotted in Fig.~\ref{fig:rates}(b). This decay lifetime exhibits a non-trivial behavior, first decreasing with increasing temperature before reaching a minimum around 100\,K and then slowly rising again up to room temperature. In order to gain deeper insights into the dynamics at play following hot carrier photoexcitation, we also plot the temperature dependence of the rise time of feature $\alpha$ in TRPL in Fig.~\ref{fig:pl_and_tals}(a), defined as the time taken for the signal to reach 90\% of its maximum value from 10\% of that value. This rise time qualitatively follows that of the 2.32-eV transient decay time over the probed temperature range. Under these excitation conditions, we thus assign the decay of this TA feature with the rate of B $\to$ A exciton interconversion. Supposing that this process is thermally activated, we fit the low temperature ($<150$\,K) decay time $\tau_{1/e}$ with an inverted Arrhenius equation which reads
\begin{equation}
    \tau_{1/e}=\left[ C \exp\left( -\frac{E_{\mathrm{act}}}{k_BT}\right)+\Gamma_0  \right]^{-1},
\end{equation}
where  $E_{\mathrm{act}}$ is the activation energy of the interconversion process, $C$ is a proportionality constant, $T$ is the temperature, $k_B$ is the Boltzmann constant and $\Gamma_0$ is the zero-temperature rate. From this, we extract an activation energy of $4 \pm 1$\,meV, which is close to the energies of low-frequency phonons observed by vibrational coherences identified on top of the TA dynamics~\cite{thouin_phonon_2019}. Those coherences involve phonons characterized by octahedral twist coupled to Pb---I---Pb bending motion, as well as Pb displacement coupled Pb---I---Pb bending. This suggests that the lattice motion coupled distinctly to excitons A and B~\cite{thouin_phonon_2019} are involved in driving the exciton interconversion dynamics.

In this section, we have exploited the apparent correlations between the rise time of feature $\alpha$ in TRPL with the decay time of feature near the A* absorption (2.32\,eV) in TA to interpret the dependence dynamics of the A* exciton on pump energy. From the TRPL measurement presented in Fig.~\ref{fig:pl_and_tals}(a), we deduced that excitons cascade from the conduction band into exciton B* and then trickle down the excitonic ladder until they reach exciton A*. Therefore, the rise time of feature $\alpha$ in TRPL corresponds to the time taken by B excitons to transfer to state A. Based on this and the matching dependencies of feature A* in TA and $\alpha$ in TRPL, we interpret the former as being related to transfer of excitons from state B to A. According to the pump energy dependence of the decay time of these features, we conjecture that the polaronic character of the excited transition dictates the transfer rate.

\subsection{Mode projection analysis}

We now examine more rigorously the mechanistic role that these lattice dynamics play in the conversion of exciton B $\to $ A. Our approach involves a simplified 
model that consists of a system with electronic states $|A\rangle$ and $|B\rangle$ 
coupled to a set $\bf Q$ of phonon modes $\bf q$ using as input experimental mode frequencies and  Huang-Rhys factors reported in reference~\citenum{thouin_phonon_2019}. 
We represent the electron-phonon coupling by the compact notation
\begin{eqnarray}
{\bf g}_{i}\cdot {\bf Q} = \sum_{q} \frac{g_{iq}}{2} \left(a_{q}^{\dagger}+a_{q}\right),
\label{coupling}
\end{eqnarray}
where ${\bf g}_{i}$ are the linear force vectors due to electrons acting on the lattice, $g_{iq}$ are coupling constants, $a_{q}^{\dagger}$ and $a_{q}$ are phonon creation and annihilation operators, respectively, and $[a_{q'},a_{q}^{\dagger}]=\delta_{qq'}$. In general, these are referenced to a particular electronic energy minimum. For the case at hand, we use the electronic ground state whereby the modes $\bf{q}$ are measured and assigned by a ground-state density-functional-theory calculation and the couplings by the experimental relative Huang-Rhys factors $\lambda_{iq}$ such that $g_{iq}^2 = (\hbar\omega_q)^2 \lambda_{iq}$~\cite{thouin_phonon_2019}. We construct a model Hamiltonian that includes the effects of electron-phonon coupling to examine the role of specific phonons in inter-exciton conversion, and we ignore many details of the 2D exciton physics such as differences in binding energies, many-body effects, etc., which are well beyond the scope of this work. Our simplifying assumption in constructing this model is that electron-phonon coupling effects dominate the nonadiabatic exciton interconversion dynamics. The model Hamiltonian that captures these effects then reads   
\begin{equation} 
 H = \left[
\begin{array}{cc}
\epsilon_{A}  & V_{ab} \\
V_{ab}^{*}   & \epsilon_{B}
\end{array}
\right]
+
\left[
\begin{array}{cc}
{\bf g}_{A}\cdot {\bf Q}&0  \\
0 & {\bf g}_{B}\cdot {\bf Q}
\end{array}
\right]  \nonumber 
+ \sum_{q}\hbar\omega_{1q}\left(a^{\dagger}_{q}a_q + \frac{1}{2}\right),
 \label{hamiltonian}
\end{equation}
where $\epsilon_{A(B)}$ is the energy of exciton A (B) and $V_{AB}$ is the phonon-mediated nonadiabatic coupling term for excitons A and B. The second term in Eq.~\ref{hamiltonian} contains the electron-phonon couplings in Eq.~\ref{coupling}, and the third term accounts for the vibrational energy and sums over all the relevant phonons $\bf q$. By diagonalizing this Hamiltonian, the role of $V_{AB}$ is to mix the diagonal states A and B, which is a necessary ingredient in the radiationless transition B $\to$ A. Physically, the spectrum of phonon-driven fluctuations in $V_{AB}$ gives rise to a so-called spectral density~\cite{nitzan2006chemical,bittner2009quantum}
\begin{eqnarray}
S_{BA}(\omega) = \int_{-\infty}^{\infty} \mathrm{d}t\,e^{-i \omega t}
C_{ BA}(t),\label{spec-dens}
\end{eqnarray}
where
\begin{align}
C_{BA}(t) =\left\langle \sum_{q}\hat V_{ABq}(t) \hat V_{BAq}(0)\right\rangle \label{cor-fun}
\end{align}
is the autocorrelation function of the polaron-transformed electron-phonon coupling operator in the Heisenberg representation, where $\langle \cdots \rangle$ denotes a thermal average. The function $C_{BA}(t)$ thus contains all the information on the nonadiabatic coupling of states A and B due to the ensemble of phonons $\bf q$. We will evaluate this function as a means to examine the role of polaronic-dressing phonons found in Ref.~\citenum{thouin_phonon_2019}. The explicit form for the kernel in Eq.~\ref{spec-dens}  is quite lengthy and is given in
Refs.~\citenum{yang2014intramolecular} and \citenum{pereverzev2006time}. 

At the adiabatic transition frequency $\tilde\omega_{BA}$, Eq.~\ref{cor-fun} produces the golden-rule B $\to$ A transition rate constant,
\begin{align}
 k_{BA}=\lim_{\tau \to \infty}2{\rm Re}\int_{0}^{\tau}\mathrm{d}t\, C_{BA}(t) e^{-i\tilde\omega_{BA}t}.
 \label{gr-expression}    
\end{align}
However, given that the absolute value of the experimental Huang-Rhys factors is unknown, our analysis does not produce the diabatic coupling (or mixing angle) needed to directly compute this rate constant, so we will restrict our analysis to the evaluation of Eq.~\ref{cor-fun}, which will permit quantitative examination of the role of specific phonons in the exciton dynamics reported above. 

In general, the sum in Eq.~\ref{cor-fun} spans all phonon modes, 
${\bf Q} = \{q_1,q_2,\cdots, q_N\}$. Insight can be
gained by analyzing which of these give the most important 
contributions to the conversion of exciton B to A.  
For this, we use the ``mode-projection'' approach 
developed by Bittner~\textit{et al}., which uses a Lanczos projection scheme to construct the smallest possible subspace 
which optimizes the electron-phonon coupling.
The result is  a list of modes ${\bf Q}_{\mathrm{opt}}
= \{q_a,q_b,q_c, \cdots \}$ ranked according to their electronic
coupling.  We can then calculate the correlation function $C_{BA}(t)$ and spectral density $S_{BA}(\omega)$
with these new modes.  Including all the modes identified in ref.~\citenum{thouin_phonon_2019} gives the exact correlation function within that subset of all possible modes. However, we gain insight into the dynamics of the transition by comparing the convergence with respect to the number of modes
included in the summation in  Eq.~\ref{cor-fun}. 
This analysis does not depend upon the diabatic coupling for Hamiltonians in which the off-diagonal term is independent of the phonon coordinates~\cite{pereverzev2009energy,Yang:2017aa,yang2014intramolecular,yang2015computing}, and therefore does not necessitate knowledge of the \emph{absolute} Huang-Rhys parameters.

 \begin{figure}[tbh]
    \centering
    \includegraphics[width=0.5\columnwidth]{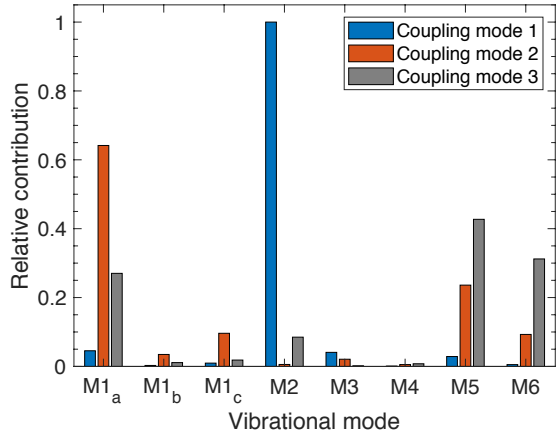}
    \caption{Electronic coupling modes projected onto the vibrationnal mode basis ranked 
    according to their contribution to the electronic coupling from exciton B to A. The labeling
    scheme is chosen to be consistent with Ref.~\citenum{thouin_phonon_2019} with the addition of subscripts to mode M1 to account for its fine structure.}
    \label{fig:erb1}
\end{figure}

In Fig.~\ref{fig:erb1}, we show the results of the 
projection analysis, showing the three dominant contributions to the electronic coupling modes (ECM) to Eq.~\ref{spec-dens}, with {\bf ECM1} $>$ {\bf ECM2} $>$ {\bf ECM3}.  For consistency, we shall refer to the 
lattice motions according to our recent work 
in Ref.~\citenum{thouin_phonon_2019} with added subscripts to mode \textbf{M1} to account for its fine structure. Briefly, \textbf{M1} and \textbf{M2} correspond to octahedral twists along axis within the inorganic lattice planes, \textbf{M3-M5} correspond to Pb---I---Pb bending modes, while \textbf{M6} corresponds to scissoring of the Pb---I---Pb bond-angle. A detailed description along with graphical representation of these modes can be found in Ref.~\citenum{thouin_phonon_2019}.

The highest-ranked mode \textbf{ECM1} is dominated by a phonon ({\bf q} = {\bf M2}) characterized by octahedral twist and Pb–I–Pb bending.  In Ref.~\citenum{thouin_phonon_2019}, we identified this as a dominant mode driving phonon coherences resulting from resonant, impulsive excitation of exciton A. The second-ranked
mode \textbf{ECM2} is dominated by a combination of
normal modes \textbf{M1} (octahedral twist along one of the two pseudocubic axes of the inorganic sheet), \textbf{M5} (Pb--–I--–Pb bending and Pb--–I stretching), and \textbf{M6} (scissoring of Pb–--I–--Pb angle), with a more modest contribution of \textbf{M2}. Finally, the third-rank mode \textbf{ECM3} is dominated by normal modes \textbf{M1}, \textbf{M5} and \textbf{M6} similar to \textbf{ECM3}. 

\begin{figure}[tbh]
    \centering
    \includegraphics[width=0.5\columnwidth]{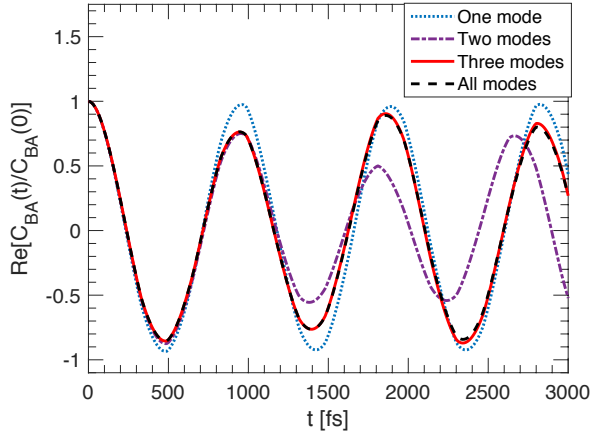}
    \caption{Real part of the normalized correlation function between exciton B and A when one, two, three or all electronic coupling modes are considered. }
    \label{fig:erb2}
\end{figure}

The correlation function (Eq.~\ref{cor-fun}) obtained by including one, two, three or all of the dominant ECMs is presented in Fig.~\ref{fig:erb2}. Over the first cycle, one mode (\textbf{ECM1}) is sufficient to reproduce the exact correlation function obtained by inclusion of all modes but quickly becomes insufficient at longer times. Inclusion of the three dominant ECMs does reproduce the exact function, and, therefore, the exact spectral density. 

On the face of this analysis, it would seem that the exciton B $\to$ A dynamics are primarily driven by the \textbf{M2} motion with other modes largely playing a spectator role by accommodating the associated energy change~\cite{pereverzev2009energy,Yang:2017aa,yang2014intramolecular,yang2015computing}. Nevertheless, the inclusion of all phonons, but predominantly \textbf{M1}, and \textbf{M5}, and \textbf{M6}, is necessary to reproduce the exact spectral density. Polaronic dressing phonons for excitons A and B therefore play active roles to a varying degree in driving inter-conversion dynamics in a way that is ranked according to Fig.~\ref{fig:erb1}.

\section*{Discussion}

In the previous section, we have reported that upon hot photocarrier injection at 5\,K, exciton B, as labelled in Fig.~\ref{fig:abs_and_pl}, is populated on a sub-picosecond timescale, and that it feeds the population of exciton A within $< 10$\,ps. We have furthermore addressed whether phonons identified via coherences resulting from impulsive stimulated Raman processes on top of the TA dynamics are spectator modes or whether they drive the observed exciton inter-conversion dynamics. We have concluded that the polaronic-dressing phonon modes drive exciton inter-conversion, with a dominant contribution of an in-plane phonon mode with frequency 35\,cm$^{-1}$, involving combined octahedral twist and Pb---I---Pb bending~\cite{thouin_stable_2018}. This conclusion highlights the role of exciton-phonon coupling, which we have previously ascribed to the polaronic nature of excitons in 2D-HOIPs~\cite{neutzner_exciton-polaron_2018}, in their relaxation dynamics. Given that the thermal activation energy for the B $\to$ A inter-conversion matches the energies of the driving phonons, and that these phonons are known to dress excitons A and B distinctly~\cite{thouin_phonon_2019}, we put forth here that polaronic modes not only dominate the vibrational coupling between excitons, but are the ones driving the inter-exciton conversion process. We underline that without knowledge of the absolute Huang-Rhys factors for each modes, it is not possible to assess their absolute contribution to the B $\to$ A excitonic inter-conversion rate as required in Eq.~\ref{gr-expression}. Nonetheless, using the mode projection analysis permits us to establish the active role of polaronic modes on exciton dynamics with confidence.  This conclusion is supported by the observation that if such an interconversion process is driven by phonons, one expects the process to have a corresponding activation energy as we observe here.

Importantly, a previous report on TRPL measurements on 2D-HOIPs at low temperature revealed similar step-like temporal evolution as the one presented in Fig.~\ref{fig:pl_and_tals}(a)~\cite{straus_direct_2016}. In that work, the spectral fine structure was interpreted as a vibronic progression from a single excitonic state. This led the authors to interpret the short-time PL decay as the non-Kasha emission of population relaxing down a unique vibrational manifold associated with the exciton. That interpretation was able to rationalize the spectral lineshape and dynamics reported in that preceeding work. While this interpretation was certainly plausible, we might expect that the excitation spectrum would be described within a Franck-Condon progression incorporating relevant optically active modes~\cite{thouin_phonon_2019}, which should be manifested throughout the full excitation profile. In Ref.~\citenum{thouin_phonon_2019}, we showed that over a phonon frequency range $\leq 50$\,cm$^{-1}$, excitons A and B show markedly different coupling to phonons, which would not be the case if they are vibronic replicas of a single exciton, where the displacement along each vibrational coordinate would be invariant. For the purposes of the analysis presented in this manuscript, we therefore conclude that each spectral feature in Fig.~\ref{fig:abs_and_pl} is a distinct exciton as we proposed in Ref.~\citenum{neutzner_exciton-polaron_2018} and concluded in Ref.~\citenum{thouin_phonon_2019}. 

Beyond the exciton nonadiabatic dynamics discussed in this work, we consider that there is broader evidence that polaronic effects determine in a more general way the exciton dynamics in 2D-HIOPs. In a separate work, we argued that polaronic effects also play an important role in modulating exciton many-body quantum dynamics~\cite{thouin_enhanced_2019}, although for different reasons as those presented in this paper. Many-body quantum dynamics are mediated by the electron-hole Coulomb field produced between excitons, which is dynamically screened by polaronic coupling. We have proposed that this polaronic protection mechanism is responsible for the weakening of exciton-exciton elastic scattering processes by two-to-three orders of magnitude compared to other monolayer two-dimensional materials that are not strongly polaronic~\cite{thouin_enhanced_2019}. Our view from that separate work is that polaronic effects are instrumental to attenuate long-range Coulomb-mediated elastic scattering. In contrast, here we argue that the phonon modes that are active in the long-range polaronic screening of excitons, in fact, are also key in the nonadiabatic electronic dynamics of the system, driving transitions from one electronic state to another. 

Furthermore, the perspective of distinct polaronic identity is supported by the observations that excitons A and B have distinct biexciton spectral structure~\cite{thouin_stable_2018}. Indeed, if distinct excitons are dressed differently by phonons, this difference should also be manifested in the resulting composite multi-particle binding energy, whereas a single exciton reflected in a vibronic progression should not display distinct binding energies by the phonon replicas.  %

The conclusions reached in this paper concerning the mechanism behind the relaxation dynamics of photoexcitations are far-reaching for the design of high-efficiency lasers and light-emitting diodes using this class of materials. Knowledge of the relaxation and recombination mechanism of injected carriers to form the emissive state is of central fundamental importance, particularly concerning the role of lattice coupling in the nonadiabatic relaxation dynamics. This report provides strong evidence of the importance of polaronic effects in this broad class of materials. Not only do they lead to a distinct dressing of photocarriers by the lattice modes~\cite{thouin_phonon_2019}, modulate exciton~\cite{neutzner_exciton-polaron_2018} and biexciton~\cite{thouin_stable_2018} binding energies, and screen them from many-body interactions~\cite{thouin_enhanced_2019}, but they also drive their relaxation dynamics. Our view is that the consequences of polaronic effects on excitonic properties of 2D-HOIPs is clear, but we have limited understanding regarding the mechanism behind this distinct polaronic nature, and indeed, the extent to which polaronic effects fundamentally give rise to multiple excitons via diverse polaron-free-carrier correlation, or whether the multiple excitons merely reflect distinct polaronic character for other reasons. This chicken-and-egg question should be central to the fundamental spectroscopic and theoretical investigation of excitons in 2D-HOIPs.

\section{Experimental Methods}

\subsection*{Sample preparation}
For the preparation of (PEA)$_2$PbI$_4$ thin films (PEA = phenylethylammonium), the precursor solution (0.25\,M) of (PEA)$_2$PbI$_4$ was prepared by mixing (PEA)I (Dyesol) with PbI$_2$ in 1:1 ratio in N,N-dimethylformamide(DMF). For example, 62.3\,mg of (PEA)I and 57.6\,mg of PbI$_2$ were dissolved in 500\,$\mu$L of DMF (anhydrous, Sigma Aldrich). The thin films were prepared by spin coating the precursor solutions on fused-silica substrates at 4000\,rpm, 30\,s, followed by annealing at 100$^{\circ}$C for 30\,min. The extensive structural characterization of these films are reported in our earlier works~\cite{thouin_stable_2018,neutzner_exciton-polaron_2018}.

\subsection*{Ultrafast differential transmission measurements}
Differential Transmission spectroscopy measurements were performed using an ultrafast laser system (Pharos Model PH1-20-0200-02-10, Light Conversion) emitting 1030-nm pulses at 100\,KHz, with an output power of 20\,W and pulse duration of $\sim 220$\,fs. Experiments were carried out in an integrated transient absorption/time-resolved photoluminescence commercial setup (Light Conversion Hera). Pump wavelengths in the spectral range 360--2600\,nm were generated by feeding 10\,W from the laser output to a commercial optical parametric amplifier (Orpheus, Light Conversion, Lithuania), while 2\,W are focused onto a sapphire crystal to obtain a single-filament white-light continuum covering the spectral range $\sim 490-1050$\,nm for the probe beam.  The probe beam transmitted through the sample was detected by an imaging spectrograph (Shamrock 193i, Andor Technology Ltd., UK)  in combination with a multichannel detector (256 pixels, 200--1100-nm wavelength sensitivity range). Energy densities used vary in the range 25--1100\,nJ/cm$^2$, most of the measurements were carried out at 215\,nJ/cm$^2$;  with a typical spot diameter of 1.9\,mm estimated at the $1/e^2$ plane). All measurements were carried in a vibration-free closed-cycle cryostation (Montana Instruments).

\subsection*{Time resolved Photoluminescence measurements} Time-resolved Photoluminescence spectroscopy was performed using tunable Ti:sapphire oscillator (Coherent Chameleon) operating at 800\,nm, 80\,MHz and an output pulsewidth of 150\,fs. The pump pulses at 400\,nm were generated by frequency doubling the fundamental output in a BBO crystal. The sample was kept in a continuous-flow static exchange gas cryostat (Oxford Instruments) to enable temperature-dependent experiments. The PL from the samples was collected and dispersed by a spectrometer (Princeton Instruments) and coupled into a Streak Camera (Hamamatsu). The maximum available time resolution is approximately 2--3\,ps.

\begin{acknowledgement}

A.R.S.K.\ acknowledges funding from EU Horizon 2020 via a Marie Sklodowska Curie Fellowship (Global) (Project No.\ 705874). F.T.\ acknowledges support from a doctoral postgraduate scholarship from the Natural Sciences and Engineering Research Council of Canada and Fond Qu\'eb\'ecois pour la Recherche: Nature et Technologies. C.S. and E.R.B. acknowledge support from the National Science Foundation (collaborative grants DMR-1904293 and DMR-1903785) C.S.\ further acknowledges support from the School of Chemistry and Biochemistry and the College of Science of Georgia Institute of Technology. 
The work at the University of Houston was also funded in
part by the  National Science Foundation (CHE-1664971, MRI-1531814) and the Robert A. Welch Foundation (E-1337).
E.R.B.\ acknowledges the Leverhulme Trust for support at 
Durham University.

\end{acknowledgement}

\begin{suppinfo}

Time-resolved photoluminescence and transient absorption decays, as indicated in the body of this article, are included as supplementary information. This material is available free of charge via the internet at \url{http://pubs.acs.org}.
\end{suppinfo}


\begin{mcitethebibliography}{29}
	\providecommand*\natexlab[1]{#1}
	\providecommand*\mciteSetBstSublistMode[1]{}
	\providecommand*\mciteSetBstMaxWidthForm[2]{}
	\providecommand*\mciteBstWouldAddEndPuncttrue
	{\def\EndOfBibitem{\unskip.}}
	\providecommand*\mciteBstWouldAddEndPunctfalse
	{\let\EndOfBibitem\relax}
	\providecommand*\mciteSetBstMidEndSepPunct[3]{}
	\providecommand*\mciteSetBstSublistLabelBeginEnd[3]{}
	\providecommand*\EndOfBibitem{}
	\mciteSetBstSublistMode{f}
	\mciteSetBstMaxWidthForm{subitem}{(\alph{mcitesubitemcount})}
	\mciteSetBstSublistLabelBeginEnd
	{\mcitemaxwidthsubitemform\space}
	{\relax}
	{\relax}
	
	\bibitem[Grancini and Nazeeruddin(2018)Grancini, and
	Nazeeruddin]{grancini2018dimensional}
	Grancini,~G.; Nazeeruddin,~M.~K. Dimensional tailoring of hybrid perovskites
	for photovoltaics. \emph{Nature Reviews Materials} \textbf{2018}, \emph{4},
	4--22\relax
	\mciteBstWouldAddEndPuncttrue
	\mciteSetBstMidEndSepPunct{\mcitedefaultmidpunct}
	{\mcitedefaultendpunct}{\mcitedefaultseppunct}\relax
	\EndOfBibitem
	\bibitem[Chen \latin{et~al.}(2018)Chen, Sun, Peng, Tang, Zheng, and
	Liang]{chen20182d}
	Chen,~Y.; Sun,~Y.; Peng,~J.; Tang,~J.; Zheng,~K.; Liang,~Z. {2D}
	{Ruddlesden-Popper} perovskites for optoelectronics. \emph{Advanced
		Materials} \textbf{2018}, \emph{30}, 1703487\relax
	\mciteBstWouldAddEndPuncttrue
	\mciteSetBstMidEndSepPunct{\mcitedefaultmidpunct}
	{\mcitedefaultendpunct}{\mcitedefaultseppunct}\relax
	\EndOfBibitem
	\bibitem[Smith \latin{et~al.}(2019)Smith, Connor, and
	Karunadasa]{smith2019tuning}
	Smith,~M.~D.; Connor,~B.~A.; Karunadasa,~H.~I. Tuning the Luminescence of
	Layered Halide Perovskites. \emph{Chemical Reviews} \textbf{2019},
	\emph{119}, 3104--3139\relax
	\mciteBstWouldAddEndPuncttrue
	\mciteSetBstMidEndSepPunct{\mcitedefaultmidpunct}
	{\mcitedefaultendpunct}{\mcitedefaultseppunct}\relax
	\EndOfBibitem
	\bibitem[Quintero-Bermudez \latin{et~al.}(2018)Quintero-Bermudez, Gold-Parker,
	Proppe, Munir, Yang, Kelley, Amassian, Toney, and
	Sargent]{quintero2018compositional}
	Quintero-Bermudez,~R.; Gold-Parker,~A.; Proppe,~A.~H.; Munir,~R.; Yang,~Z.;
	Kelley,~S.~O.; Amassian,~A.; Toney,~M.~F.; Sargent,~E. Compositional and
	orientational control in metal halide perovskites of reduced dimensionality.
	\emph{Nature Materials} \textbf{2018}, \emph{17}, 900\relax
	\mciteBstWouldAddEndPuncttrue
	\mciteSetBstMidEndSepPunct{\mcitedefaultmidpunct}
	{\mcitedefaultendpunct}{\mcitedefaultseppunct}\relax
	\EndOfBibitem
	\bibitem[Blancon \latin{et~al.}(2018)Blancon, Stier, Tsai, Nie, Stoumpos,
	Traoré, Pedesseau, Kepenekian, Katsutani, Noe, Kono, Tretiak, Crooker,
	Katan, Kanatzidis, Crochet, Even, and Mohite]{blancon_scaling_2018}
	Blancon,~J.-C. \latin{et~al.}  Scaling law for excitons in {2D} perovskite
	quantum wells. \emph{Nature Communications} \textbf{2018}, \emph{9},
	2254\relax
	\mciteBstWouldAddEndPuncttrue
	\mciteSetBstMidEndSepPunct{\mcitedefaultmidpunct}
	{\mcitedefaultendpunct}{\mcitedefaultseppunct}\relax
	\EndOfBibitem
	\bibitem[Hong \latin{et~al.}(1992)Hong, Ishihara, and Nurmikko]{Hong1992}
	Hong,~X.; Ishihara,~T.; Nurmikko,~A. Dielectric confinement effect on excitons
	in {PbI}$_4$-based layered semiconductors. \emph{Physical Review B}
	\textbf{1992}, \emph{45}, 6961--6964\relax
	\mciteBstWouldAddEndPuncttrue
	\mciteSetBstMidEndSepPunct{\mcitedefaultmidpunct}
	{\mcitedefaultendpunct}{\mcitedefaultseppunct}\relax
	\EndOfBibitem
	\bibitem[Ishihara \latin{et~al.}(1992)Ishihara, Hong, and Ding]{Ishihara1992}
	Ishihara,~T.; Hong,~X.; Ding,~J. Dielectric confinement effect for exciton and
	biexciton states in {Pbla}-based two-dimensional semiconductor structures.
	\emph{Surface Science} \textbf{1992}, \emph{267}, 323--326\relax
	\mciteBstWouldAddEndPuncttrue
	\mciteSetBstMidEndSepPunct{\mcitedefaultmidpunct}
	{\mcitedefaultendpunct}{\mcitedefaultseppunct}\relax
	\EndOfBibitem
	\bibitem[Thouin \latin{et~al.}(2018)Thouin, Neutzner, Cortecchia, Dragomir,
	Soci, Salim, Lam, Leonelli, Petrozza, Kandada, and Silva]{thouin_stable_2018}
	Thouin,~F.; Neutzner,~S.; Cortecchia,~D.; Dragomir,~V.~A.; Soci,~C.; Salim,~T.;
	Lam,~Y.~M.; Leonelli,~R.; Petrozza,~A.; Kandada,~A. R.~S.; Silva,~C. Stable
	biexcitons in two-dimensional metal-halide perovskites with strong dynamic
	lattice disorder. \emph{Physical Review Materials} \textbf{2018}, \emph{2},
	034001\relax
	\mciteBstWouldAddEndPuncttrue
	\mciteSetBstMidEndSepPunct{\mcitedefaultmidpunct}
	{\mcitedefaultendpunct}{\mcitedefaultseppunct}\relax
	\EndOfBibitem
	\bibitem[Neutzner \latin{et~al.}(2018)Neutzner, Thouin, Cortecchia, Petrozza,
	Silva, and Srimath~Kandada]{neutzner_exciton-polaron_2018}
	Neutzner,~S.; Thouin,~F.; Cortecchia,~D.; Petrozza,~A.; Silva,~C.;
	Srimath~Kandada,~A.~R. Exciton-polaron spectral structures in two-dimensional
	hybrid lead-halide perovskites. \emph{Physical Review Materials}
	\textbf{2018}, \emph{2}, 064605\relax
	\mciteBstWouldAddEndPuncttrue
	\mciteSetBstMidEndSepPunct{\mcitedefaultmidpunct}
	{\mcitedefaultendpunct}{\mcitedefaultseppunct}\relax
	\EndOfBibitem
	\bibitem[Even \latin{et~al.}(2014)Even, Pedesseau, and
	Katan]{even2014understanding}
	Even,~J.; Pedesseau,~L.; Katan,~C. Understanding quantum confinement of charge
	carriers in layered {2D} hybrid perovskites. \emph{ChemPhysChem}
	\textbf{2014}, \emph{15}, 3733--3741\relax
	\mciteBstWouldAddEndPuncttrue
	\mciteSetBstMidEndSepPunct{\mcitedefaultmidpunct}
	{\mcitedefaultendpunct}{\mcitedefaultseppunct}\relax
	\EndOfBibitem
	\bibitem[Zhang \latin{et~al.}(2018)Zhang, Fang, Tokina, Long, and
	Prezhdo]{zhang2018rapid}
	Zhang,~Z.; Fang,~W.-H.; Tokina,~M.~V.; Long,~R.; Prezhdo,~O.~V. Rapid
	decoherence suppresses charge recombination in multi-layer 2D halide
	perovskites: Time-domain ab initio analysis. \emph{Nano Letters}
	\textbf{2018}, \emph{18}, 2459--2466\relax
	\mciteBstWouldAddEndPuncttrue
	\mciteSetBstMidEndSepPunct{\mcitedefaultmidpunct}
	{\mcitedefaultendpunct}{\mcitedefaultseppunct}\relax
	\EndOfBibitem
	\bibitem[Kataoka \latin{et~al.}(1993)Kataoka, Kondo, Ito, Sasaki, Uchida, and
	Miura]{kataoka_magneto-optical_1993}
	Kataoka,~T.; Kondo,~T.; Ito,~R.; Sasaki,~S.; Uchida,~K.; Miura,~N.
	Magneto-optical study on excitonic spectra in
	({C}$_6${H}$_{13}${NH}$_3$)$_2${PbI}$_4$. \emph{Physical Review B}
	\textbf{1993}, \emph{47}, 2010--2018\relax
	\mciteBstWouldAddEndPuncttrue
	\mciteSetBstMidEndSepPunct{\mcitedefaultmidpunct}
	{\mcitedefaultendpunct}{\mcitedefaultseppunct}\relax
	\EndOfBibitem
	\bibitem[Thouin \latin{et~al.}(2019)Thouin, Valverde-Chávez, Quarti,
	Cortecchia, Bargigia, Beljonne, Petrozza, Silva, and
	Srimath~Kandada]{thouin_phonon_2019}
	Thouin,~F.; Valverde-Chávez,~D.~A.; Quarti,~C.; Cortecchia,~D.; Bargigia,~I.;
	Beljonne,~D.; Petrozza,~A.; Silva,~C.; Srimath~Kandada,~A.~R. Phonon
	coherences reveal the polaronic character of excitons in two-dimensional lead
	halide perovskites. \emph{Nature Materials} \textbf{2019}, \emph{18},
	349--356\relax
	\mciteBstWouldAddEndPuncttrue
	\mciteSetBstMidEndSepPunct{\mcitedefaultmidpunct}
	{\mcitedefaultendpunct}{\mcitedefaultseppunct}\relax
	\EndOfBibitem
	\bibitem[Pereverzev \latin{et~al.}(2009)Pereverzev, Bittner, and
	Burghardt]{pereverzev2009energy}
	Pereverzev,~A.; Bittner,~E.~R.; Burghardt,~I. Energy and charge-transfer
	dynamics using projected modes. \emph{The Journal of Chemical Physics}
	\textbf{2009}, \emph{131}, 034104\relax
	\mciteBstWouldAddEndPuncttrue
	\mciteSetBstMidEndSepPunct{\mcitedefaultmidpunct}
	{\mcitedefaultendpunct}{\mcitedefaultseppunct}\relax
	\EndOfBibitem
	\bibitem[Yang and Bittner(2014)Yang, and Bittner]{yang2014intramolecular}
	Yang,~X.; Bittner,~E.~R. Intramolecular Charge- and Energy-Transfer Rates with
	Reduced Modes: {C}omparison to {M}arcus Theory for Donor--Bridge--Acceptor
	Systems. \emph{The Journal of Physical Chemistry A} \textbf{2014},
	\emph{118}, 5196--5203\relax
	\mciteBstWouldAddEndPuncttrue
	\mciteSetBstMidEndSepPunct{\mcitedefaultmidpunct}
	{\mcitedefaultendpunct}{\mcitedefaultseppunct}\relax
	\EndOfBibitem
	\bibitem[Yang and Bittner(2015)Yang, and Bittner]{yang2015computing}
	Yang,~X.; Bittner,~E.~R. Computing intramolecular charge and energy transfer
	rates using optimal modes. \emph{The Journal of Chemical Physics}
	\textbf{2015}, \emph{142}, 244114\relax
	\mciteBstWouldAddEndPuncttrue
	\mciteSetBstMidEndSepPunct{\mcitedefaultmidpunct}
	{\mcitedefaultendpunct}{\mcitedefaultseppunct}\relax
	\EndOfBibitem
	\bibitem[Yang \latin{et~al.}(2017)Yang, Keane, Delor, Meijer, Weinstein, and
	Bittner]{Yang:2017aa}
	Yang,~X.; Keane,~T.; Delor,~M.; Meijer,~A. J. H.~M.; Weinstein,~J.;
	Bittner,~E.~R. Identifying electron transfer coordinates in
	donor-bridge-acceptor systems using mode projection analysis. \emph{Nature
		Communications} \textbf{2017}, \emph{8}, 14554\relax
	\mciteBstWouldAddEndPuncttrue
	\mciteSetBstMidEndSepPunct{\mcitedefaultmidpunct}
	{\mcitedefaultendpunct}{\mcitedefaultseppunct}\relax
	\EndOfBibitem
	\bibitem[Straus \latin{et~al.}(2016)Straus, Hurtado~Parra, Iotov, Gebhardt,
	Rappe, Subotnik, Kikkawa, and Kagan]{straus_direct_2016}
	Straus,~D.~B.; Hurtado~Parra,~S.; Iotov,~N.; Gebhardt,~J.; Rappe,~A.~M.;
	Subotnik,~J.~E.; Kikkawa,~J.~M.; Kagan,~C.~R. Direct {Observation} of
	{Electron}–{Phonon} {Coupling} and {Slow} {Vibrational} {Relaxation} in
	{Organic}–{Inorganic} {Hybrid} {Perovskites}. \emph{Journal of the American
		Chemical Society} \textbf{2016}, \emph{138}, 13798--13801\relax
	\mciteBstWouldAddEndPuncttrue
	\mciteSetBstMidEndSepPunct{\mcitedefaultmidpunct}
	{\mcitedefaultendpunct}{\mcitedefaultseppunct}\relax
	\EndOfBibitem
	\bibitem[Wu \latin{et~al.}(2015)Wu, Trinh, Niesner, Zhu, Norman, Owen, Yaffe,
	Kudisch, and Zhu]{wu2015trap}
	Wu,~X.; Trinh,~M.~T.; Niesner,~D.; Zhu,~H.; Norman,~Z.; Owen,~J.~S.; Yaffe,~O.;
	Kudisch,~B.~J.; Zhu,~X.-Y. Trap states in lead iodide perovskites.
	\emph{Journal of the American Chemical Society} \textbf{2015}, \emph{137},
	2089--2096\relax
	\mciteBstWouldAddEndPuncttrue
	\mciteSetBstMidEndSepPunct{\mcitedefaultmidpunct}
	{\mcitedefaultendpunct}{\mcitedefaultseppunct}\relax
	\EndOfBibitem
	\bibitem[Giovanni \latin{et~al.}(2018)Giovanni, Chong, Liu, Dewi, Yin, Lekina,
	Shen, Mathews, Gan, and Sum]{giovanni2018coherent}
	Giovanni,~D.; Chong,~W.~K.; Liu,~Y. Y.~F.; Dewi,~H.~A.; Yin,~T.; Lekina,~Y.;
	Shen,~Z.~X.; Mathews,~N.; Gan,~C.~K.; Sum,~T.~C. Coherent Spin and
	Quasiparticle Dynamics in Solution-Processed Layered {2D} Lead Halide
	Perovskites. \emph{Advanced Science} \textbf{2018}, \emph{5}, 1800664\relax
	\mciteBstWouldAddEndPuncttrue
	\mciteSetBstMidEndSepPunct{\mcitedefaultmidpunct}
	{\mcitedefaultendpunct}{\mcitedefaultseppunct}\relax
	\EndOfBibitem
	\bibitem[Schmitt-Rink \latin{et~al.}(1985)Schmitt-Rink, Chemla, and
	Miller]{schmitt1985theory}
	Schmitt-Rink,~S.; Chemla,~D.; Miller,~D. Theory of transient excitonic optical
	nonlinearities in semiconductor quantum-well structures. \emph{Physical
		Review B} \textbf{1985}, \emph{32}, 6601\relax
	\mciteBstWouldAddEndPuncttrue
	\mciteSetBstMidEndSepPunct{\mcitedefaultmidpunct}
	{\mcitedefaultendpunct}{\mcitedefaultseppunct}\relax
	\EndOfBibitem
	\bibitem[Price \latin{et~al.}(2015)Price, Butkus, Jellicoe, Sadhanala, Briane,
	Halpert, Broch, Hodgkiss, Friend, and Deschler]{price2015hot}
	Price,~M.~B.; Butkus,~J.; Jellicoe,~T.~C.; Sadhanala,~A.; Briane,~A.;
	Halpert,~J.~E.; Broch,~K.; Hodgkiss,~J.~M.; Friend,~R.~H.; Deschler,~F.
	Hot-carrier cooling and photoinduced refractive index changes in
	organic--inorganic lead halide perovskites. \emph{Nature Communications}
	\textbf{2015}, \emph{6}, 8420\relax
	\mciteBstWouldAddEndPuncttrue
	\mciteSetBstMidEndSepPunct{\mcitedefaultmidpunct}
	{\mcitedefaultendpunct}{\mcitedefaultseppunct}\relax
	\EndOfBibitem
	\bibitem[Giovanni \latin{et~al.}(2016)Giovanni, Chong, Dewi, Thirumal, Neogi,
	Ramesh, Mhaisalkar, Mathews, and Sum]{giovanni_tunable_2016}
	Giovanni,~D.; Chong,~W.~K.; Dewi,~H.~A.; Thirumal,~K.; Neogi,~I.; Ramesh,~R.;
	Mhaisalkar,~S.; Mathews,~N.; Sum,~T.~C. Tunable room-temperature
	spin-selective optical {Stark} effect in solution-processed layered halide
	perovskites. \emph{Science Advances} \textbf{2016}, \emph{2}, e1600477\relax
	\mciteBstWouldAddEndPuncttrue
	\mciteSetBstMidEndSepPunct{\mcitedefaultmidpunct}
	{\mcitedefaultendpunct}{\mcitedefaultseppunct}\relax
	\EndOfBibitem
	\bibitem[Hunsche \latin{et~al.}(1993)Hunsche, Heesel, Ewertz, Kurz, and
	Collet]{hunsche1993spectral}
	Hunsche,~S.; Heesel,~H.; Ewertz,~A.; Kurz,~H.; Collet,~J. Spectral-hole burning
	and carrier thermalization in GaAs at room temperature. \emph{Physical Review
		B} \textbf{1993}, \emph{48}, 17818\relax
	\mciteBstWouldAddEndPuncttrue
	\mciteSetBstMidEndSepPunct{\mcitedefaultmidpunct}
	{\mcitedefaultendpunct}{\mcitedefaultseppunct}\relax
	\EndOfBibitem
	\bibitem[Nitzan(2006)]{nitzan2006chemical}
	Nitzan,~A. \emph{Chemical dynamics in condensed phases: relaxation, transfer
		and reactions in condensed molecular systems}; Oxford university press,
	2006\relax
	\mciteBstWouldAddEndPuncttrue
	\mciteSetBstMidEndSepPunct{\mcitedefaultmidpunct}
	{\mcitedefaultendpunct}{\mcitedefaultseppunct}\relax
	\EndOfBibitem
	\bibitem[Bittner(2009)]{bittner2009quantum}
	Bittner,~E.~R. \emph{Quantum dynamics: applications in biological and materials
		systems}; CRC press, 2009\relax
	\mciteBstWouldAddEndPuncttrue
	\mciteSetBstMidEndSepPunct{\mcitedefaultmidpunct}
	{\mcitedefaultendpunct}{\mcitedefaultseppunct}\relax
	\EndOfBibitem
	\bibitem[Pereverzev and Bittner(2006)Pereverzev, and
	Bittner]{pereverzev2006time}
	Pereverzev,~A.; Bittner,~E.~R. Time-convolutionless master equation for
	mesoscopic electron-phonon systems. \emph{The Journal of Chemical Physics}
	\textbf{2006}, \emph{125}, 104906\relax
	\mciteBstWouldAddEndPuncttrue
	\mciteSetBstMidEndSepPunct{\mcitedefaultmidpunct}
	{\mcitedefaultendpunct}{\mcitedefaultseppunct}\relax
	\EndOfBibitem
	\bibitem[Thouin \latin{et~al.}(2019)Thouin, Cortecchia, Petrozza, Kandada, and
	Silva]{thouin_enhanced_2019}
	Thouin,~F.; Cortecchia,~D.; Petrozza,~A.; Kandada,~A. R.~S.; Silva,~C.
	\emph{arXiv:1904.12402 [cond-mat, physics:physics]}, arXiv:1904.12402
	[cond-mat.mtrl-sci]\relax
	\mciteBstWouldAddEndPuncttrue
	\mciteSetBstMidEndSepPunct{\mcitedefaultmidpunct}
	{\mcitedefaultendpunct}{\mcitedefaultseppunct}\relax
	\EndOfBibitem
\end{mcitethebibliography}
\providecommand{\latin}[1]{#1}
\makeatletter
\providecommand{\doi}
{\begingroup\let\do\@makeother\dospecials
	\catcode`\{=1 \catcode`\}=2 \doi@aux}
\providecommand{\doi@aux}[1]{\endgroup\texttt{#1}}
\makeatother
\providecommand*\mcitethebibliography{\thebibliography}
\csname @ifundefined\endcsname{endmcitethebibliography}
{\let\endmcitethebibliography\endthebibliography}{}

\end{document}